\documentclass[11pt]{article}
\usepackage{graphicx,floatflt,amssymb} 
\usepackage{epsfig} 
\usepackage{graphics} 
\usepackage{psfrag}
\usepackage{axodraw} 
\input epsf
\usepackage{epsfig,axodraw}

\def\beq{\begin{equation}}
\def\eeq{\end{equation}}
\def\bea{\begin{eqnarray}}
\def\eea{\end{eqnarray}}
\def\bq{\begin{quote}}
\def\eq{\end{quote}}
\def \lsim{\mathrel{\vcenter
     {\hbox{$<$}\nointerlineskip\hbox{$\sim$}}}}
\def \gsim{\mathrel{\vcenter
     {\hbox{$>$}\nointerlineskip\hbox{$\sim$}}}}
\def\gappeq{\mathrel{\rlap {\raise.5ex\hbox{$>$}}
{\lower.5ex\hbox{$\sim$}}}}
\def\lappeq{\mathrel{\rlap{\raise.5ex\hbox{$<$}}
{\lower.5ex\hbox{$\sim$}}}}

\def\bea{\begin{eqnarray}}   
\def\eea{\end{eqnarray}}

\begin{document}
\vspace*{-1in}
\renewcommand{\thefootnote}{\fnsymbol{footnote}}
\begin{flushright}
\texttt{CI-UAN/02-15T}\\
\texttt{LPT-Orsay/03-40}\\
\texttt{hep-ph/yymmddd} 
\end{flushright}
\vskip 5pt
\begin{center}
{\Large {\bf Leptogenesis with four gauge singlets
}}
\vskip 25pt
{\bf Asmaa Abada $^{1,}$\footnote{E-mail address:
abada@lyre.th.u-psud.fr}
and Marta Losada $^{2,}$}\footnote{E-mail address:
malosada@uan.edu.co} 
 
\vskip 10pt  
$^1${\it Laboratoire de Physique Th\'eorique, 
Universit\'e de Paris XI, B\^atiment 210, 91405 Orsay Cedex,
France} \\
$^2${\it Centro de Investigaciones, 
Universidad Antonio Nari\~{n}o, Cll. 58A No. 37-94, Santa Fe de Bogot\'{a},
Colombia}
\vskip 20pt
\vskip 20pt {\bf Abstract}
\end{center}

\begin{quotation} {\noindent
We consider a generic type of leptogenesis model which can successfully produce
the correct value of the observed baryon number to entropy ratio.  The main
feature of this model is that it is a simple  TeV scale model, a scale
accessible in near future machines,  with a minimal particle content. Both
supersymmetric and non-supersymmetric  versions of the model are feasible. 
This model  also gives left-handed neutrino masses compatible  with all current
data from direct and indirect neutrino experiments.
\vskip 10pt 
\noindent PACS number(s):~12.60.Jv, 14.60.Pq, 11.30.Fs\\
}

\end{quotation} \noindent{Leptogenesis, Neutrino Physics, Baryon Asymmetry of
the Universe}

\vskip 20pt  

\setcounter{footnote}{0} \renewcommand{\thefootnote}{\arabic{footnote}}

 \newpage
\section{Introduction}

The determination of a mechanism which can solve the problem of the baryon
asymmetry of the Universe (BAU) continues to be a great challenge  for both
particle physics and cosmology. Among the different mechanisms there has been
renewed interest in  leptogenesis \cite{FY} due to  current experimental
results from neutrino (astro)physics which could  be a first indication of
violation of lepton number if neutrinos have Majorana masses.  In contrast
there has been no direct experimental evidence for baryon number violation in
elementary particle interactions. However, recent experimental data from WMAP
\cite{WMAP} has determined the best fit values of the baryon  density
$\Omega_b h^2$  and the ratio of the baryon  to  photon density $\eta =
\frac{n_B}{n_{\gamma}}$,

\bea \Omega_b h^2 &=& 0.0224 \pm 0.0009 \nonumber\\ \eta &=&
6.5^{+0.4}_{-0.3}\times 10^{-10}. \eea

We now summarize the main ingredients of the leptogenesis mechanism which 
could generate the BAU. Without a primordial asymmetry, as usual  it is
necessary that Sakharov's  conditions be fulfilled \cite{sakharov}, that is:
violation of Baryon Number, violation of C and CP and a deviation from thermal
equilibrium. These requirements are generic for any baryogenesis or
leptogenesis mechanism.  The leptogenesis scenario relies on two separate
instances: first the production of a lepton asymmetry and secondly the
conversion of this asymmetry into a produced baryon asymmetry. The lepton
asymmetry could be produced for instance in the out-of-equilibrium decay of a
particle with lepton number and CP violating interactions.  It has been argued
that the natural scale for this to happen is from $10^{10}-10^{15}$ GeV. This
energy scale  is not directly testable making it very difficult to verify a
given model. The second stage occurs via sphaleron interactions which are not
suppressed for temperatures, \cite{kuzmin}, $10^2<T<10^{12}$ GeV. These
interactions occur in  weak interaction theories in which all (left) chiral
fermions participate without conserving B nor L.

Furthermore, different leptogenesis models  have tried to  simultaneously give
an adequate explanation for non-zero values of neutrino masses, normally via
the see-saw mechanism \cite{yanagida2,Ramond}, with a common interaction being
at the origin of the asymmetry  and mass generation.

Many of the models that have been considered are based on the original Fukugita
and Yanagida model \cite{FY}  including three right-handed neutrinos
\cite{3RHLepto} in addition to the Standard Model. It is the decay of the right
handed neutrinos that produces the lepton asymmetry~\footnote{Several models of
this type are imbedded into (SUSY) GUT models such as
SO(10)\cite{Plumacher}.}.  An alternative model \cite{masarkar} originally
devised in the context of the Left-Right Symmetric Model produces the lepton
asymmetry  via the decay of a triplet Higgs boson field. It is worth mentioning
that one of the difficulties that occurs in the construction of a viable model
of leptogenesis when the heavy decaying particle is not a gauge singlet arises
due to the possible washout processes. To avoid any elimination of the
asymmetry it is necessary that {\it all} interactions in which the decaying
particle is involved should be  out of equilibrium.

Recent work by Buchmuller, Di Bari and Plumacher  \cite{bdp1,bdp2,bdp3}  have
studied in a lot of  detail the case of the SM + 3 right-handed neutrinos
solving the corresponding Boltzmann equations that take into account the effect
of  washout processes for  temperatures below the mass scale of the lightest
right-handed neutrino as well as the thermal production of the initial
abundance of right-handed neutrinos. A lower bound on the mass on the
right-handed neutrino can be placed $M \gsim 10^8 $GeV \cite{bdp1,di}. One of
the most interesting features of the latest leptogenesis analysis has been the
determination of an upper bound on the sum of the light neutrino masses $\sim
0.2$ eV,  which is stronger than the  WMAP one \cite{WMAP}  $\sim 0.7$ eV, by
requiring the appropriate amount of baryon asymmetry \cite{bdp2,bdp3}. 

The inclusion of right-handed (RH) neutrinos directly leads to the possibility
of giving  small  masses to the light left-handed  neutrinos via the see-saw
mechanism \cite{yanagida2,Ramond}. Since  the RH neutrinos are singlets under
the Standard Model gauge group, it is possible  to include Majorana mass terms
for these particles. These terms in conjunction with  the Yukawa terms in the 
Lagrangian,
coupling the left-handed with  right-handed neutrinos via the Higgs 
boson interaction,   produce the necessary structure to
induce a see-saw mass term for the left-handed  neutrinos. It is  a  simple 
way to explain  the left-handed neutrinos mass scale.

It is interesting to consider an alternative energy scale, $\sim$ TeV,  at
which the lepton asymmetry could be produced and which additionally could be
testable at future experiments. Several of the difficulties that arise in this
scenario have been discussed in reference \cite{hambye}. These include: a)
obtaining adequate values of the lepton asymmetry given  the constraint from
the out-of-equilibrium decay,  b) damping effects if the decaying particle has
gauge interactions,  c) a much too small value of the neutrino masses, or  d)
it is difficult to have a model in which the same interactions produce the
asymmetry and non-zero neutrino masses.  The author of Ref.~\cite{hambye} also
discusses possible enhancement mechanisms which could overcome the above
mentioned difficulties. Other possible TeV scale models  have been recently
discussed in Refs. \cite{Senami,Senami2,Boubekeur}.

In this paper we present a (generic type of) TeV scale model which  can be an
interesting candidate for a viable leptogenesis model. In section 2 we present 
the model and its main features. In section 2.1 we present a non
supersymmetric  toy model with only two generations  to point out the main
characteristics of the model. In section 2.2 we illustrate possible textures
which can be used successfully in the case with four gauge singlets. 
In section
2.3 we comment on additional features of the supersymmetric version of the
model. We finally conclude in section 3.

\section{The Model}

Let us first explore a simple (non-supersymmetric) version of the model to
distinguish certain features. The Lagrangian is given by\footnote{We do not
write out explicitly the terms involving quark fields.},

\beq L = L_{SM} + \bar{\psi}_{R_{I}} i \partial \! \! \! /~\psi_{R_{I}} -
\frac{M_{N_{I}}}{2}(\bar{\psi}_{R_{I}}^{c} \psi_{R_{I}} + h.c.) - 
(Y^{\nu}_{IJ} \bar{L}_{J} \psi_{R_{I}} \phi + h.c.) \ ,\label{lag} \eeq where $
\psi_{R_{I}}$ are two-component spinors describing the right-handed neutrinos 
and we define a  Majorana 4-component spinor, $N_{I} =  \psi_{R_{I}} + 
\psi_{R_{I}}^{c}$. Our index I runs from 0 to 3. The zeroth component of $L_I$
corresponds to a left-handed lepton doublet which must satisfy the LEP
constraints from the Z- width on a fourth left-handed neutrino~\cite{pdg}. The 
$Y_{IJ}$ are Yukawa couplings and  the field $\phi$ is the SM Higgs boson
doublet whose vacuum expectation value is denoted by $v_u$.

We work in the basis in which the mass matrix for the right-handed neutrinos
$M$  is diagonal,

\beq M = diag(M_0,M_1,M_2,M_3) \eeq and define $m_D = Y_{\nu} v_u$. The
neutrino mass matrix for the  left-handed neutrinos is given by,

\beq m_{\nu}  = m_D^T M^{-1} m_D = Y_{\nu}^{\dagger} M^{-1} Y_{\nu} v_u^2\ .
\eeq which is diagonalized by the matrix $U$. We will consider the
out-of-equilibrium decay of the  lightest of the gauge singlets $N_I$, which we
take to be  $N_1$.  The decay rate at tree-level 
 is given by

\beq \Gamma_{N_{I}} = \Gamma(N_I \rightarrow L_J + \phi^*) +  \Gamma(N_I
\rightarrow L_J^* + \phi) =  \frac{1}{(8 \pi)} [Y_{\nu} Y_{\nu}^{\dagger}]_{II}
M_{N_{I}} . \eeq

To ensure an out-of equilibrium decay of $N_1$  it is necessary that
$\frac{\Gamma_{N_{I}}}{H(T=M_{1})} \ll 1$, where H is the Hubble  expansion
rate at $T=M_{1}$. 
This
condition can also be expressed in terms of the quantity $\tilde{m}_1$,  an
effective mass parameter  \cite{bp} defined as

\beq \tilde{m}_1 =   \frac{(m_D^{\dagger} m_D)_{11}}{M_1}  \lsim 5 \times
10^{-3} ~{\mathrm eV}. \eeq This effective mass $\tilde{m}_1$ depends on $g_{*}$
, which is  the effective
number of relativistic degrees of freedom at $T=M_1$,  for four generations 
the upper bound  will slightly increase.

\begin{figure}[htb]
\unitlength1mm
\SetScale{2.8}
\begin{boldmath}
\begin{center}
\begin{picture}(60,30)(0,-10)
\Line(0,0)(20,0)
\Line(45,15)(20,0)
\DashLine(45,-15)(20,0){2}
\Text(-2,0)[r]{$N_I$}
\Text(45,17)[l]{$\bar{L}_J$}
\Text(45,-17)[c]{$\phi$}
\end{picture}
\hspace{1cm}
\begin{picture}(60,30)(0,-10)
\Line(0,0)(20,0)
\Line(45,15)(20,0)
\DashLine(45,-15)(20,0){2}
\Text(-2,0)[r]{$N_I$}
\Text(45,17)[l]{$L_J$}
\Text(45,-17)[c]{$\bar{\phi}$}
\end{picture}
\end{center}
\end{boldmath}
\caption{Decay Diagrams at Tree-level.}
\protect\label{tree}
\end{figure}
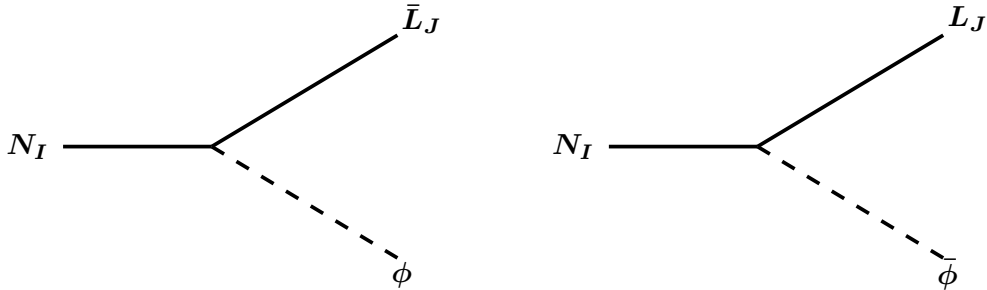

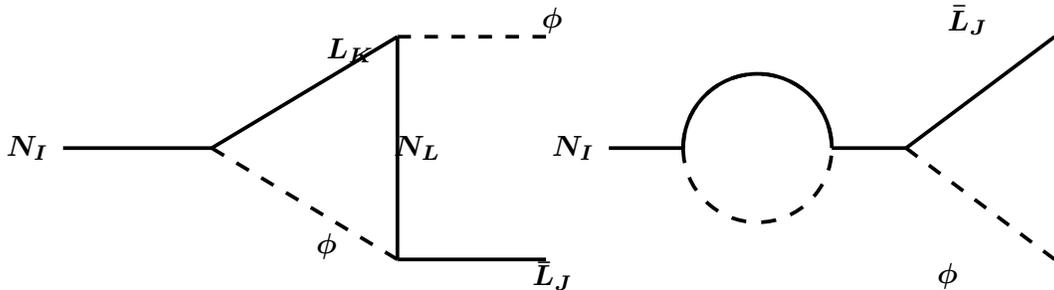
\begin{figure}[htb]
\unitlength1mm
\SetScale{2.8}
\begin{boldmath}
\begin{center}
\begin{picture}(60,30)(0,-10)
\Line(0,0)(20,0)
\Line(45,15)(20,0)
\DashLine(45,-15)(20,0){2}
\Line(45,15)(45,-15)
\DashLine(45,15)(65,15){2}
\Line(45,-15)(65,-15)
\Text(-2,0)[r]{$N_I$}
\Text(35,13)[l]{$L_K$}
\Text(35,-13)[c]{$\phi$}
\Text(47,0)[c]{$N_{L}$}
\Text(65,17)[c]{$\phi$}
\Text(65,-17)[c]{$\bar{L}_J$}
\end{picture}
\hspace{1cm}
\begin{picture}(60,30)(0,-10)
\Line(0,0)(10,0)
\CArc(20,0)(10,0,180)
\DashCArc(20,0)(10,-180,180){2}
\Line(30,0)(40,0)
\Line(60,15)(40,0)
\DashLine(60,-15)(40,0){2}
\Text(-2,0)[r]{$N_I$}
\Text(45,17)[l]{$\bar{L}_J$}
\Text(45,-17)[c]{$\phi$}
\end{picture}
\end{center}
\end{boldmath}
\caption{One-loop diagrams contributing to the decay
  $\Gamma(N_I\rightarrow \bar{L}_J \phi)$. There
are similar diagrams that contribute to  $\Gamma(N_I\rightarrow L_J
\bar{\phi})$ .}
\protect\label{oneloop}
\end{figure}

 The CP asymmetry calculated from the
interference of the tree diagrams of figure \ref{tree} with the one-loop
diagrams (self-energy and vertex corrections) given in figure \ref{oneloop} is
\cite{roulet}

\beq \epsilon_I = \frac{1}{(8\pi)}\frac{1}{  [Y_{\nu} Y_{\nu}^{\dagger}]_{II}}
\sum_J {\mathrm Im}{ [Y_{\nu}
Y_{\nu}^{\dagger}]^2_{IJ}}\left[f\left(\frac{M_{J}^2}{M_{I}^{2}}\right) + g
\left(\frac{M_{J}^2}{M_{I}^{2}}\right)\right] \label{asym}\ , \eeq where

\bea f(x) &=& \sqrt x[1 - (1+x)\ln\frac{1+x}{x}], \nonumber\\ g(x) &=&
\frac{\sqrt x}{1-x}. \eea 

The full lepton asymmetry $\eta_L$ will be given by,

\beq \eta_L = \frac{1}{f}\frac{n_{\nu_{R}}}{n_{\gamma}}\epsilon d\ , 
\eeq where  $\epsilon$ is the CP asymmetry, $f$ is the dilution factor from photon production from the time of leptogenesis to recombination\footnote{ For our numerical estimates we take $f=10^{2}$.},  d is  the washout factor which
takes into account  dilution effects of inverse decays, scatterings,
annihilations, etc., which affect the final result of the lepton number density
and  $\frac{n_{\nu_{R}}}{n_{\gamma}}$ determines the
initial abundance of the right-handed neutrinos and depends on  the production
mechanism. In order to obtain the exact value of the  latter two quantities, 
one must solve the corresponding Boltzmann equations. In this paper we will not
solve the  Boltzmann equations but will work under the assumption of having an
inital thermal abundance for the right-handed  neutrinos, unless explicitly
stated, and we will check that the values of our Yukawa couplings are such that
the washout  effects will be small and not affect the overall features of the
model. In an upcoming paper we will solve  in detail the Boltzmann equations
for this model \cite{am3}.

The ratio $\eta$  of the  baryon to photon density  is related to $\eta_L$ by 

\beq	\eta= \frac{n_B}{n_{\gamma}} = - \left(\frac{8N_{F} + 4N_{H}}{22N_F +
13N_H}\right)  \eta_{L}, \eeq where $N_H$ is the number of Higgs doublets and
$N_F$ is the number  of fermionic families.  

Using the fact that the matrix $\Omega = \frac{v_u}{(m_I M_J)^{1/2}}
U^{\dagger} Y_{\nu}$ is orthogonal \cite{casasibarra} and following the
procedure of Ref.~\cite{di} it is easy to show that the upper bound of the
CP-asymmetry produced in the decay of the lightest right-handed neutrino $N_1$
is now \beq |\epsilon_1| \lsim \frac{3}{8\pi} \frac{M_1 m_4}{v_u^2},
\label{cpbound} \eeq where $m_4$ denotes the largest eigenvalue of the
left-handed neutrino mass matrix  $m_{\nu}$. Due to experimental constraints we
require that $m_4 > 45$~GeV, which implies that the bound on $\epsilon_1$ is
irrelevant. We will have regions of parameter space in which the produced CP
asymmetry can be very large, thus the final allowed region  of parameter space
will include areas in which the washout processes can be very large as well.

\subsection{Toy Model}

As a simple toy model consider the  Lagrangian
of eq. (\ref{lag})  for the case with only two generations; that is index $I$
runs from $0, 1$. 
 The
decay rate and CP asymmetry are given by,

\beq \Gamma_1 = \frac{1}{8\pi} (y_{11}y_{11}^{*} + y_{10} y_{10}^{*}) M_1 \ ,
\eeq

\beq \epsilon_{1} = \frac{1}{8\pi} \left( f(M_0^2/M_1^2) + 
g(M_0^2/M_1^2)\right) \frac{{\mathrm Im}[ y_{10}^{*} y_{10}^{*}y_{00}y_{00} + 2
y_{10}^{*}y_{11}^{*} y_{00}y_{01} + y_{11}^{*}y_{11}^{*}
y_{01}y_{00}]}{y_{11}y_{11}^{*} + y_{10} y_{10}^{*}}\ . \label{epsilon1} \eeq

\begin{figure}[htbp]

\psfrag{y}{$y$} \psfrag{i}{$i$} \psfrag{x}{$\times$}
\hspace{2cm}\includegraphics[scale=0.70]{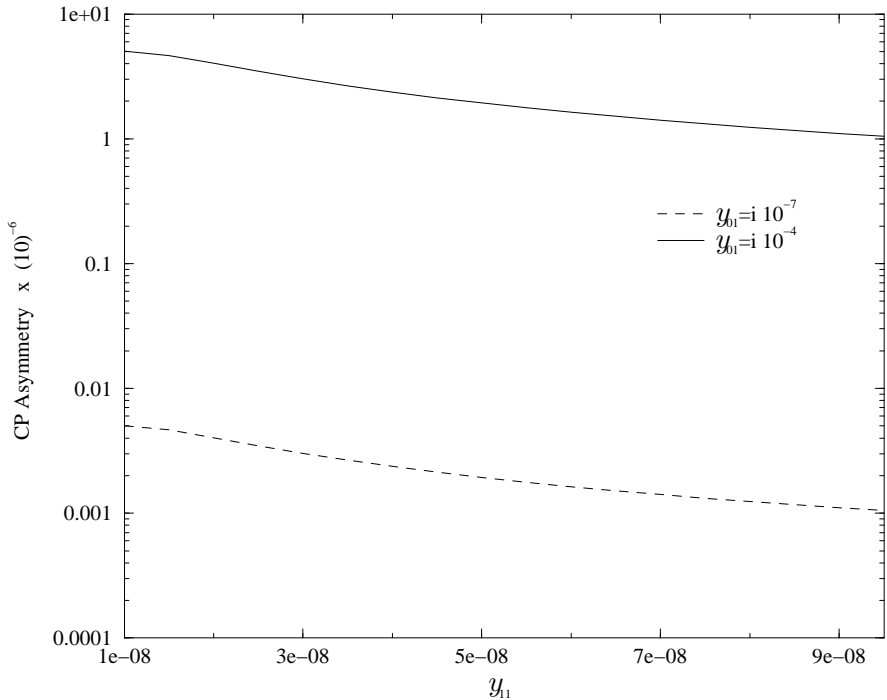} 
 \caption{$\epsilon_1$ as a function of
$y_{11}$ for the first texture T$_1$ considered in the text. The solid line
corresponds to $y_{01}= i 10^{-4}$, the dashed line is for $y_{01}= i 10^{-7}$.
The rest of the parameters are defined in the text.}

\protect\label{cpasym}

\end{figure}

 We will analyze two possible textures for the Yukawa coupling matrix.  We
first will consider a texture which has a straightforward implementation when
we consider the 4 generations case, see below. The second texture allows us to
illustrate the interplay of the different terms which contribute to the CP
asymmetry in eq. (\ref{cpasym}). Both textures will be useful when the
supersymmetric version of the model is considered. The first texture T$_1$ is
of the form

\beq Y_{\nu} = \left(\begin{array}{cc} \epsilon & \epsilon\\ \alpha & 1\\
\end{array}\right). \eeq This just means that $y_{11} \sim y_{10} \sim
\epsilon$ and $y_{01} \sim \alpha$ while $y_{00} \sim 1$. We choose to insert
the CP phase of the Yukawa couplings in $\alpha$~\footnote{It is possible to
parametrize the CP violation of the neutrino sector \`a la Jarlskog. However,
we will leave this issue for future work. }. Thus,  the first term in the
numerator of the expression for $\epsilon_1$ in eq. (\ref{epsilon1}) does not
contribute to the asymmetry.

\begin{figure}[htbp]

\hspace{3cm}

\psfrag{y}{$y$}
\psfrag{i}{$i$} \psfrag{mm}{$\tilde m_1$}
\hspace{2cm}\includegraphics[scale=0.70]{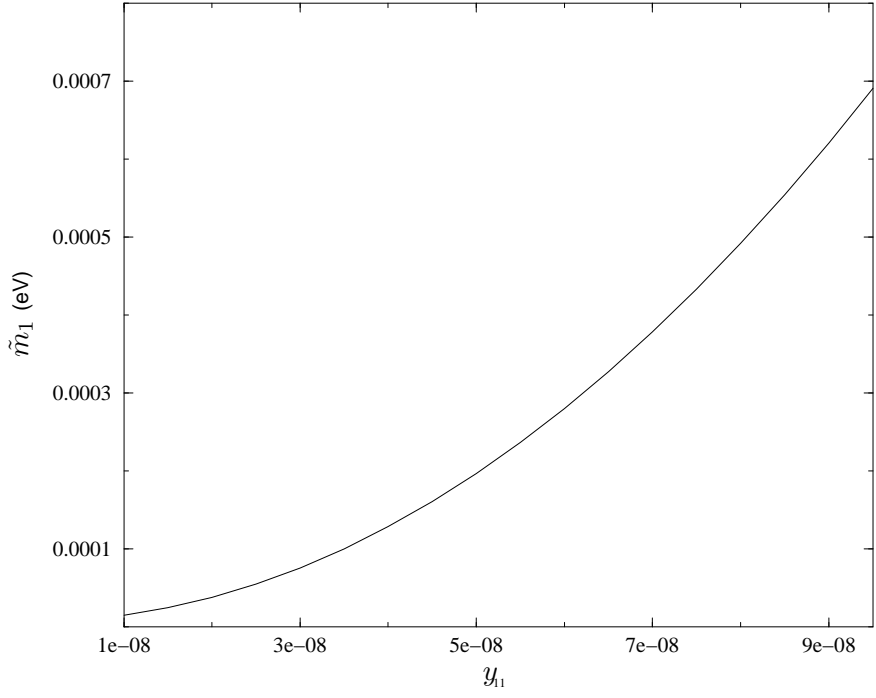}  \caption{The effective
mass $\tilde{m}_1$ as a function of $y_{11}$ for the first texture T$_1$. The
rest of the parameters are defined in the text.}

\protect\label{m1tilde}

\end{figure}

 In figures \ref{cpasym} and \ref{m1tilde} we plot the value of  $\epsilon_1$
and $\tilde{m}_1$ as a function of the coupling $y_{11}$ for $M_1=400$GeV, 
$M_0 = 650$ GeV, $y_{00} = 1, y_{10}= 10^{-8}$ and for two values of $ y_{01}= 
i 10^{-4}, i 10^{-7}$. We are taking these values for the sake of illustration,
as they  allow us to satisfy all constraints from low energy data and
cosmology. For the same values of the parameters  we obtain a heavy left-handed
neutrino with a mass above $45 $ GeV, and a light neutrino with a  mass on the
order of $10^{-4}$ eV.

We can then see
quite easily that for these values of the Yukawa  couplings $\Gamma_{N_{1}}/H
\ll 1$, (or equivalently $\tilde{m}_1 <10^{-3}$ eV) which  allows for an out-of
equilibrium decay of $N_1$ and suppresses the washout  from inverse decays,
$\Delta L=2$ and $\Delta L=1$ scatterings. The decay rate and the decay
temperature increase as $y_{11}, y_{10}$ increase making it harder to have an
out-of equilibrium decay and increasing the suppression of the final lepton
asymmetry.

\begin{figure}[htbp]

\hspace{3cm}

\psfrag{y}{$y$}
\psfrag{i}{$i$} \psfrag{mm}{$\tilde m_1$ (eV)} \psfrag{eta}{$\eta$}
\hspace{2cm}\includegraphics[scale=0.70]{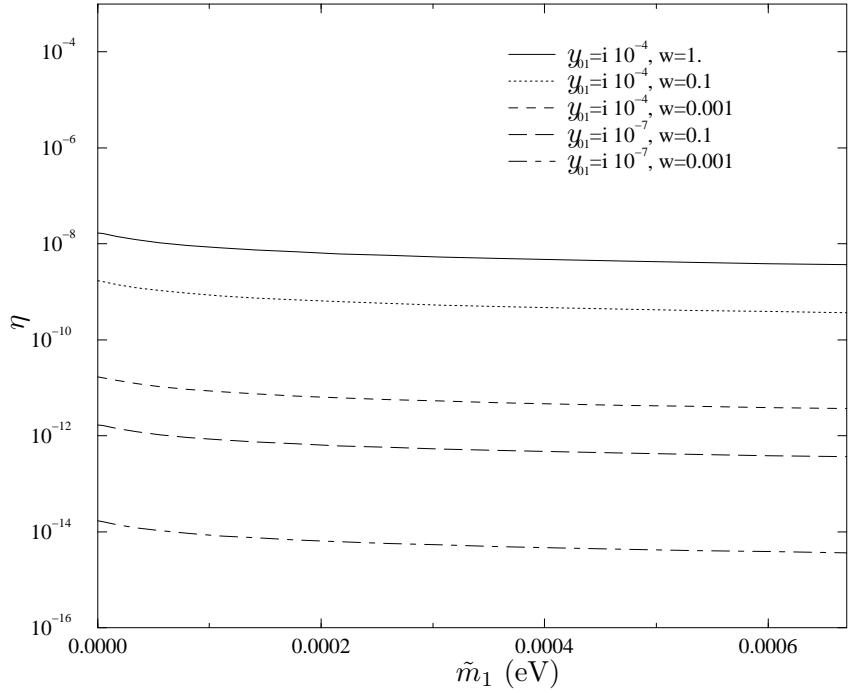}  \caption{The baryon
asymmetry $\eta$ as a function of  $\tilde{m}_1$ in the case of the first
texture T$_1$ for different values of the coupling $y_{01}$ and the product 
w$=\frac{n_{\nu_{R}}}{n_{\gamma}} d$. The rest of the
parameters are defined  in the text.}

\protect\label{bau}

\end{figure}

In figure \ref{bau} we plot the baryon asymmetry $\eta$ as a function of
 $\tilde{m}_1$ for two values of $ y_{01}=  i 10^{-4}, i 10^{-7}$ and  for
 three different values of the product of  $\frac{n_{\nu_{R}}}{n_{\gamma}} d = 1, 0.1, 0.001$.


\begin{figure}   [htbp]

 \centering    
\hspace{-1cm}  \begin{tabular}{cc}      
\includegraphics[scale=0.50]{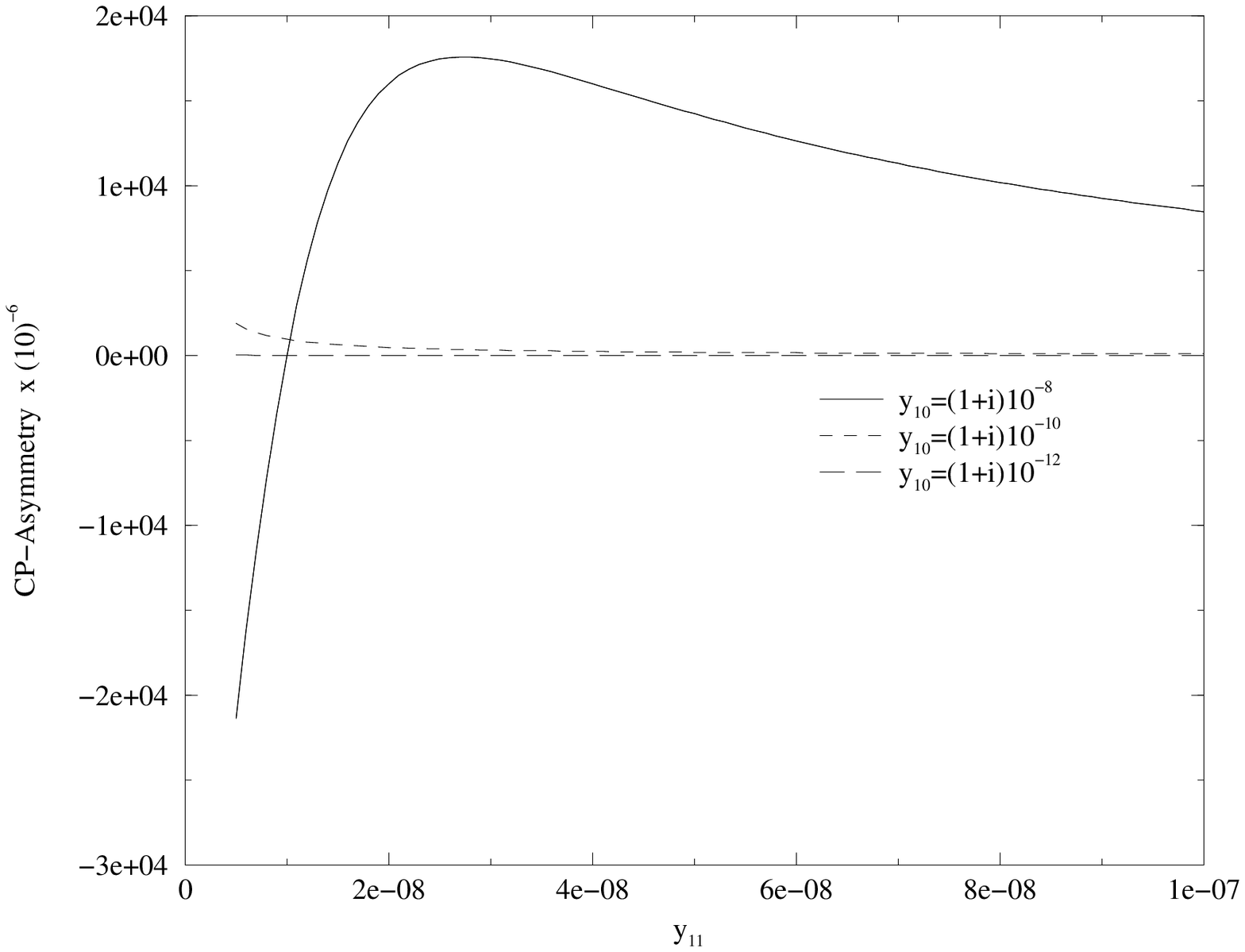} &     
\includegraphics[scale=0.40]{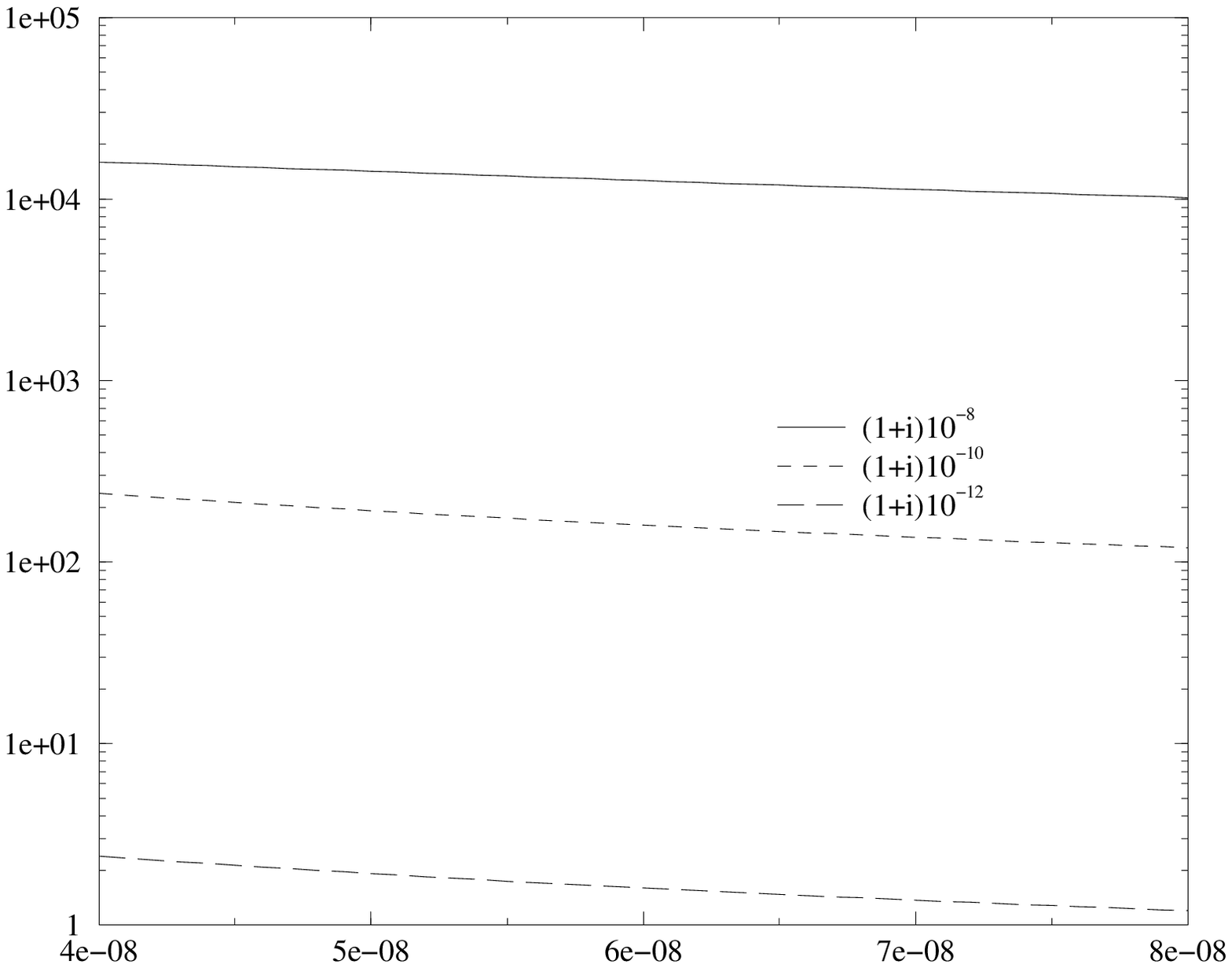}     
  \\     
\end{tabular}     
 
 \caption{$\epsilon_1$ as a function of
$y_{11}$ for the second texture T$_2$ considered in the text. The solid line
corresponds to $y_{10}= (1+i) 10^{-8}$, the dashed line is for  $y_{10} = (1
+i) 10^{-10}$ and the long-dashed line to  $y_{10} = (1 +i) 10^{-12}$. The
rest of the parameters are  defined in the text. The left part is a zoom to
separate the $y_{10} = (1+i) 10^{-10}$ and $y_{10} = (1+i) 10^{-12}$ lines.}
       \protect\label{cpasy2}  
\end{figure}

\begin{figure}[htbp]

\psfrag{y}{$y$}
\psfrag{i}{$i$} \psfrag{w}{$\tilde m_1$} \psfrag{mm}{$\tilde m_1$}
\hspace{2cm}\includegraphics[scale=0.70]{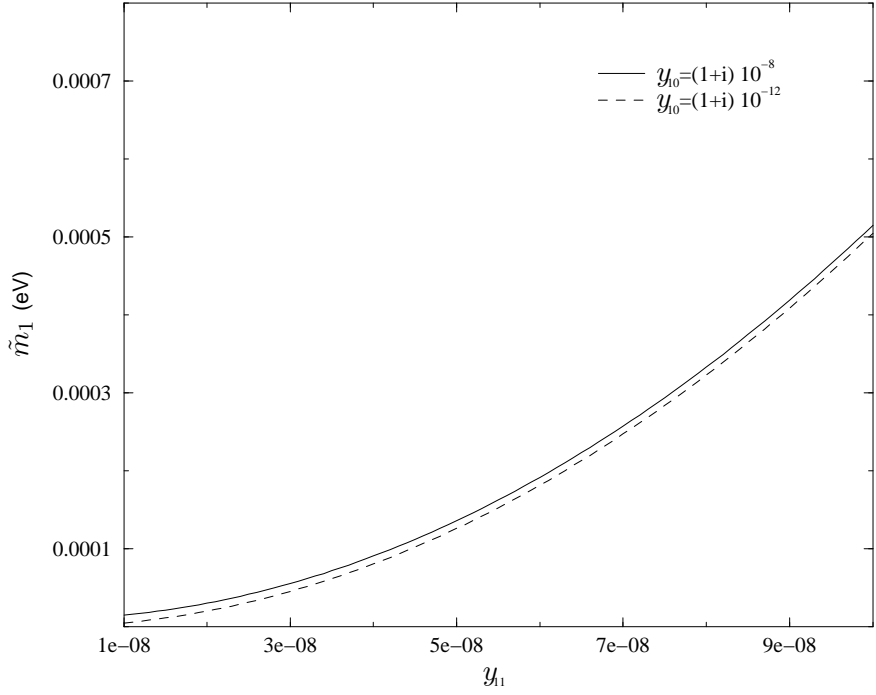}  
\caption{The effective
mass $\tilde{m}_1$ as a function of $y_{11}$ for the second texture T$_2$. The
solid line corresponds to $y_{10}= (1+i) 10^{-8}$ and  the dashed line is for 
$y_{10} = (1 +i) 10^{-12}$. The rest of the parameters are defined in the
text.}

\protect\label{m1tilde2}

\end{figure}

\begin{figure}   [htbp]

 \centering    
\hspace{-1cm}  \begin{tabular}{cc}      
\psfrag{i}{$i$} \psfrag{w}{$\tilde m_1$} \psfrag{mm}{$\tilde m_1$ (eV)}
\psfrag{eta}{$\eta$}
\includegraphics[scale=0.50]{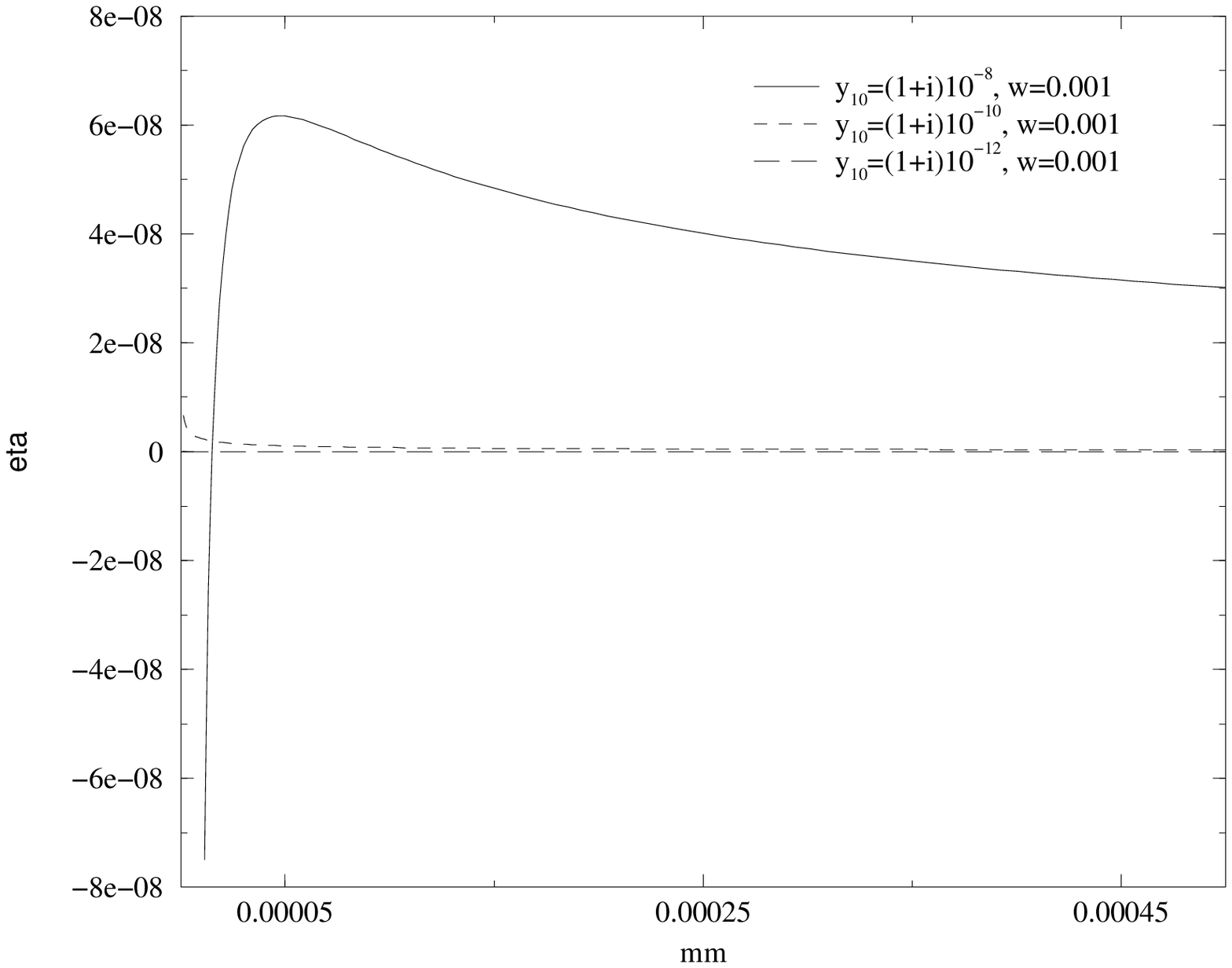} &     
\includegraphics[scale=0.40]{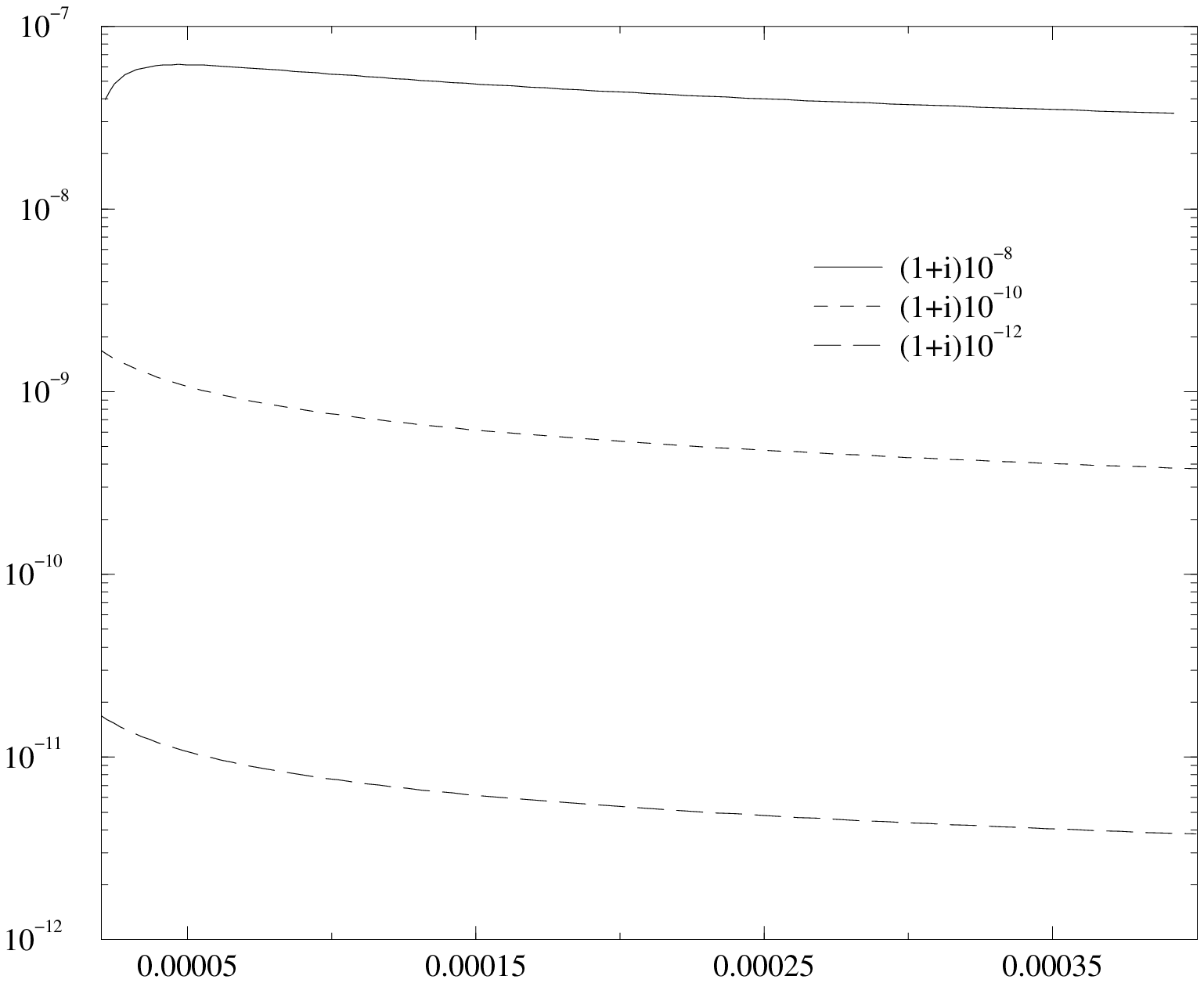}     
  \\     
\end{tabular}  \caption{The baryon asymmetry $\eta$ as a function of  $\tilde{m}_1$ in the
case of the second texture T$_2$ for three different values of the coupling 
$y_{10}$ and for w$=\frac{n_{\nu_{R}}}{n_{\gamma}}
d=0.001$. The coupling $y_{11}$ varies from $10^{-9}$ to $10^{-7}$ and the rest
of the parameters are defined  in the text.The left part is a zoom to
separate the $y_{10} = (1+i) 10^{-10}$ and $y_{10} = (1+i) 10^{-12}$ lines.}

\protect\label{bau2}

\end{figure}

The second texture T$_2$ we would like to analyze has a Yukawa coupling matrix
of the form \beq Y_{\nu} = \left(\begin{array}{cc} \epsilon &  \alpha\\ 1 & 
1\\ \end{array}\right). \eeq 
Once again we choose to include the CP phase in
the Yukawa couplings only  in $\alpha$   which allows us to see the interplay
of different terms contributing to the CP asymmetry in eq. (\ref{epsilon1}). 
In
this case we can have larger values of the right-handed neutrino masses $M_1$
and $M_0$,  and we can still obtain appropiate values for  neutrinos masses. We
plot in figure \ref{cpasy2} the values of $\epsilon_1$ as a function of
$y_{11}$ for $y_{00}=1, y_{01} = 1, M_{1} = 600$ GeV, $ M_0 = 1$ TeV for $
y_{10} = (1 +i) 10^{-8}, (1+i) 10^{-10}, (1+i)10^{-12}$. The value of the CP
asymmetry can be quite large and in order to obtain an adequate value of
$\epsilon_1$ we must impose a hierarchy between the couplings which enter the
decay rate, $y_{11}$ and $y_{10}$\footnote{We can also choose values of the
couplings that enter the decay rate such that $\tilde{m}_1 > 10^{-3}$eV which
implies that washout processes can be large and it is necessary to solve the
Boltzmann equations to obtain a precise value for the final baryon asymmetry.}.
In figure \ref{m1tilde2} we plot the value of $\tilde{m}_1$ as a function of
the coupling $y_{11}$ for two values of $ y_{10}= (1+ i) 10^{-8}, (1+ i)
10^{-12}$, the other parameters are taken to be the same as the ones chosen to
plot $\epsilon_1$ ($y_{00}=1, y_{01} = 1, M_{1} = 600$ GeV, $ M_0 = 1$ TeV).
Figure \ref{bau2} shows the value of the baryon asymmetry $\eta$ as a function of
$\tilde{m}_1$ for   $\frac{n_{\nu_{R}}}{n_{\gamma}} d = 
0.001$. 

\begin{figure}[htbp]

\hspace{3cm}

 \psfrag{y}{$y$}
\psfrag{i}{$i$} \hspace{2cm}\includegraphics[scale=0.70]{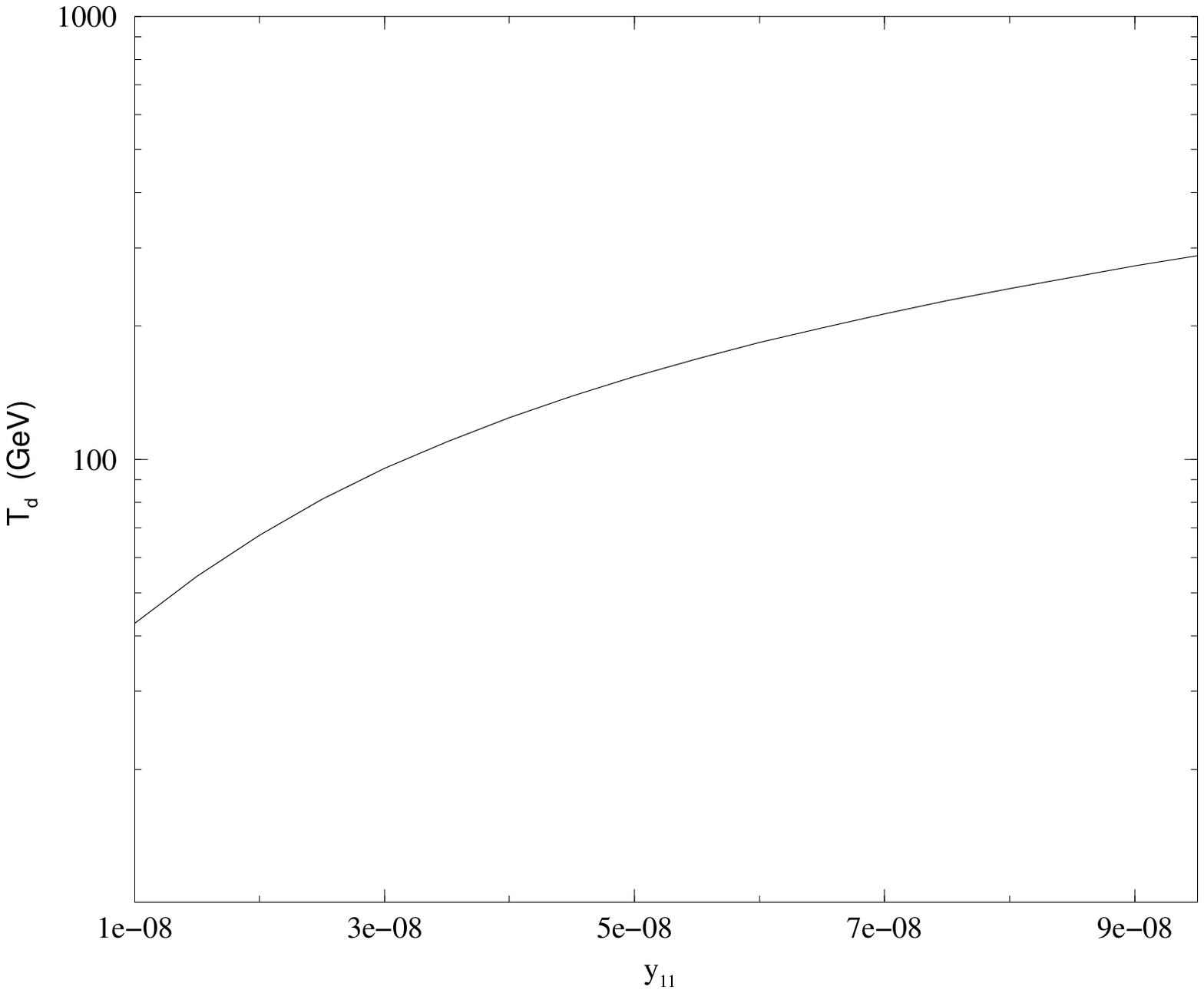} 
\caption{The decay temperature $T_d$ as a function of $y_{11}$ for the first
texture T$_1$, for $M_1=400$GeV,  $M_0 = 650$ GeV, $y_{00} = 1$ 
and $y_{10}=10^{-8}$.}

\protect\label{decayT}

\end{figure}

In this toy model an additional aspect that must be taken into
account is the  decay temperature for the RH neutrinos, as it can be just above
the electroweak phase transition  temperature. The reason for this is that in
order to obtain adequate values for the LH  neutrino masses, $M_1$ and $M_0$
cannot be of the order of  (1-10) TeV. This will not be an important issue in
the supersymmetric version of this toy model. In figure \ref{decayT} we  plot
the decay temperature $T_d$ as a function of $y_{11}$ for the set of 
parameters described in the caption.

There are a few important issues that must be discussed. At first sight it
would seem that we are creating an asymmetry on both light ($L_1$) and heavy
left-handed leptons ($L_0$). However, as has been shown in refs.
\cite{Plumacher,bdp1,bp} there are an important class of washout processes
which are proportional to $m_i^2$, where $m_i$ denotes  the eigenmasses of the
LH neutrino mass matrix. For the $L_0$ fermion, this mass is large  ($> 45
$GeV). Thus this type of processes will washout any asymmetry in $L_0$ and we
remain only with the asymmetry in $L_1$.

An important point in the analysis of Ref.~\cite{bdp3} is that for values 
of $\tilde{m}_1 <10^{-3}$~eV,  washout processes will
not affect significantly the value of the baryon asymmetry.
 For our model this
will continue to be true as well, however, as mentioned above, in our case we
will have regions where the initial asymmetry can be very large as well as regions
where the dilution effects are large or the initial abundance of the RH neutrinos is
small which will still remain viable. We also note that as in the usual model of the
SM + 3 right-handed neutrinos $\tilde{m}_1 > m_1$, where $m_1$ denotes the
smallest eigenvalues of $m_{\nu}$.

The main conclusion for our model is that there clearly  is enough room to
produce the BAU at the TeV scale even if there is a strong suppression of
$\eta$ from processes which occur at $T<M_1$.

\subsection{4 Generations}

One of the possible textures for $Y_{\nu}$ in the case of four generations of
leptons which can generate the appropiate amount of  CP-asymmetry and fit
neutrino data is of the form

\beq Y_{\nu} = C \left(\begin{array}{cccc} \epsilon & \epsilon & \epsilon &
\epsilon\\ \epsilon & 1 & 1 & 0 \\ \epsilon & 1 & 1 & 0 \\ \alpha & 0 & 0 &
1/C\\ \end{array}\right). \eeq This will induce to first order a mass matrix
for the light left-handed neutrinos  of

\beq m_{\nu} =  C^2 \frac{v_u^2}{M}\left(\begin{array}{ccc} \epsilon^2 &
\epsilon & \epsilon\\ \epsilon & 1 & 1\\ \epsilon & 1 & 1 \end{array}\right),
\eeq which is a simple form of the light neutrino mass matrix which can account
for all data \cite{yanagida,Ramond2}. $C$ is a small number that makes $C^2
v_u^2/M$ to be of the correct order of magnitude.

\subsection{Supersymmetric Version of the Model}

The supersymmetrized version of our toy model is very straightforward. The
additional contributions to the usual MSSM superpotential are~\footnote{Terms of
the form $S_{IJK} N_I N_J N_K$ are neglected we will consider the possible
implications of the inclusion of this term in particular regarding the
possibility of generating a spontaneous vacuum expectation value for the scalar
component of the singlet superfields in an upcoming paper \cite{ams2}.}

\beq W = \frac{1}{2} Y_{eIJ}^k L^I L^J E^c_k + Y_{\nu IJ} L^J H_u N^{c I} + 
M_I N^I N^I. \eeq The zeroth component of $L_I$ is now the down-type Higgsino
and there is no full fourth generation of fermions. The decay rate of $N_I$ is given
by,

\beq \Gamma_{N_{I}}= \Gamma(N_I\rightarrow L_J H_u) +  \Gamma(N_I\rightarrow
\tilde{L}_{J} \tilde{H}_{u}) =  \frac{1}{(8 \pi)} [Y_{\nu}
Y_{\nu}^{\dagger}]_{II} M_{N_{I}}. \eeq There is an analogous expression for
the CP asymmetry produced in the decay of $N_1$ to that of equation
(\ref{asym}). There are however, a few crucial points:

\begin{itemize}

\item  In this case  there are new Yukawa couplings for  Higgsino-gauge singlet
fields which enter into the expression for $\epsilon_1$ which are not
constrained to be tiny by left-handed neutrino mass bounds. The upper bound on
the value of the asymmetry, analogous to eq. (\ref{cpbound}), is now
proportional to the neutral down-type Higgsino mass. \item Due to the existence
of the $\mu$-term which couples the $H_u$ and $H_d$  fields, the asymmetry in
the $H_d$ will be washed out leaving only the asymmetry in the $L_i$ fields
when the lightest gauge singlet decays, $i=1,2,3$.

\item A clear distinction with the non-supersymmetric case is that here we are
not required to have such a low value for $M_1$, as the Higgsino has other
contributions to its mass arising from the $\mu$-term. So now we can easily
have values for $M_1 \sim (1-10)$ TeV and still obtain an adequate value for
the asymmetry and neutrino masses and mixings.

\item The right-handed neutrino mass spectrum can be much more hierarchical.

\item  For a Yukawa matrix $Y_{\nu}$ with a texture of the second type
presented in section 2.1 we can then deviate from having both $y_{00} \sim
y_{01}\sim 1$ and the hierarchy between $y_{11}$ and $y_{10}$ and still obtain
an adequate value of the baryon asymmetry as we are less constrained by
neutrino data.

\item The gravitino problem which appears in supersymmetric leptogenesis models
with 3 RH neutrinos is avoided in our model. \end{itemize}

\section{Conclusions and Outlook}

We have presented a new and simple model that can provide an adequate scenario
for leptogenesis at the TeV scale. The inclusion of an additional gauge singlet
allows new contributions to the CP asymmetry which are not constrained to be 
small. Thus it is relatively simple to obtain the correct amount of baryon
asymmetry. The detailed discussion of the toy model has allowed us to
illustrate essential features of the model. Neutrino masses and mixings that
satisfy low energy experimental constraints can also be obtained. Our
preliminary numerical analysis has been  very simple,  we will address the
important issues involved in solving the Boltzmann equations in an upcoming
paper\cite{am3}.

The supersymmetric version of our model is also quite interesting as it could
generate dynamically the $\mu$ terms of the superpotential as well as the
masses for the right-handed neutrinos. We will explore these aspects in more
detail in a further analysis \cite{ams2}.

\section*{Acknowledgements}

We would like to thank S. Davidson for extremely useful discussions.  The work
of M.L. was partially supported by Colciencias-BID, under contract no.
120-2000. We thank CERN Theory Division for hospitality during the completion
of this work.

\end{document}